\documentstyle[sprocl,amsbsy,amssymb]{article}
\title{Issues in Gravitational Wave Data Analysis}
\author{Lee Samuel Finn}
\address{Physics and Astronomy, Northwestern University, Evanston IL
  60208-3112, USA}
\begin{document}
\maketitle 

\abstracts{Data analysis is the application of probability and
  statistics to draw inference from observation.  Is a signal present
  or absent?  Is the source an inspiraling binary system or a
  supernova?  At what point in the sky is the radiation incident from?
  In these notes I want to address how two different statistical
  methodologies --- Bayesian and Frequentist --- approach the problem
  of statistical inference.}

\section{Introduction}

There is a perception that Bayesian and Frequentist statistical
methodologies are, at root, identical; that attempts to distinguish
between them are sophistry; that, even if there are differences, they
are only semantic and without any operational consequence.

These are all serious misconceptions. Bayesian and Frequentist
statistical methodologies are inequivalent; they ask fundamentally
different questions of the data in an attempt to draw inferences; and,
because they ask different questions, the analyses undertaken are
quantitatively and qualitatively different and lead to conclusions
different in type and kind, even when regarding the same data set.

In \S\ref{sec:Bayesian} and \S\ref{sec:Frequentist} we pose the 
question ``with what confidence can we conclude that, in the last 
hour, the gravitational waves from a new core collapse supernova in 
the Virgo cluster of galaxies passed through our gravitational wave 
detector?''  In \S\ref{sec:Bayesian} we consider the question from a 
Bayesian perspective, while in \S\ref{sec:Frequentist} we consider it 
from a Frequentist perspective.  We will find that this apparently 
straightforward question takes on a different meaning to the 
Frequentist and the Bayesian, leading each to respond in different 
ways.

Finally, in \S\ref{sec:example} we examine an example of a
multi-detector data analysis problem --- the detection of a stochastic
gravitational wave signal in two detectors.  A Frequentist analysis of
this problem has been
developed~\cite{michelson87a,christensen92a,flanagan93a,allen97c} and
is summarized first.\footnote{See also Allen's contribution to this
  proceedings.} In the following subsection we explore a Bayesian
analysis of the same problem~\cite{finn97b}, which leads to an analysis
that is entirely unlike the Frequentist one.  Together, the discussion
in \S\S2--4 demonstrates the very real, operational differences
between the Bayesian and Frequentist approaches to data analysis.

\section{Learning From Observation: Bayesian Data
  Analysis}\label{sec:Bayesian} 

With what confidence can we conclude that, in the last hour, the
gravitational waves from a new core collapse supernova in the Virgo
cluster of galaxies passed through our gravitational wave detector?

We don't approach this, or any other question, without some prior 
expectations.  In this case, before examining the observations, our 
prior understanding of astrophysics leads us to expect, on average, 
one such core collapse every 4 months; consequently, we believe the 
probability is approximately $3.4\times10^{-4}$ that in any given hour 
--- including the last --- gravitational waves from a new Virgo 
cluster supernova were incident on our detector.

Probability, as we have used it here, means {\em degree of belief.} In 
this instance, our degree of belief coincided with the {\em expected 
frequency\/} of supernova events; however, this need not be the case: 
we can assess degree of belief even when we can't assess relative 
frequency.  For example, suppose that I have a coin that is known to 
be heavily biased toward either heads or tails.  What is your degree 
of belief that, when I next flip the coin, it will land heads-up?  
Without telling you the direction of the bias, you can't evaluate the 
expected relative frequency of heads or tails.  You can, however, 
quantify your degree of belief: having no more reason to believe that 
the bias is toward heads than towards tails, you have no more reason 
to believe that the coin will, when next flipped, land heads-up than 
that it will land heads-down.  Your degree of belief in either 
alternative, then, is $1/2$.

One does not have to search either long or hard to find examples from
astrophysics where probability as ``degree of belief'' exists and
probability as ``expected frequency'' does not.  For example, what is
the probability that there exists a cosmological stochastic
gravitational wave signal with a given amplitude and spectrum?  In this
case, ``expected frequency'' has no meaning: there is only one
Universe, and it either does or does not have a stochastic
gravitational wave background of given spectrum and amplitude.

After we examine the output of our gravitational-wave detector, our
degree of belief in the supernova proposition may change: we may, on
the basis of the observations, become more or less certain that
radiation from a supernova passed through our detector.  How do
observations change our degree of belief in the different
alternatives?

To explore how our degree of belief evolves with the examination of 
observations we need to introduce some notation:
\begin{eqnarray}
H_{0} &=& \left(\begin{array}{l}
\mbox{proposition that gravitational waves from a}\\
\mbox{new supernova in the Virgo cluster {\em did not}}\\
\mbox{pass through our detector in the last hour}
\end{array}
\right),\\
{\cal I}&=& \left(\begin{array}{l}
\mbox{our prior knowledge of astrophysics, including}\\
\mbox{our best assessment of the supernova rate}
\end{array}\right),\\
g &=& \left(\begin{array}{l}
\mbox{observations from our gravitational wave detector}
\end{array}\right),\\
P(A|B) &=& \left(\begin{array}{l}
    \mbox{degree of belief in $A$ assuming that $B$ is true}
\end{array}\right),\\
\neg{}A &=& \left(\mbox{logical negation of proposition $A$}\right).
\end{eqnarray}
In this notation, $P(H_{0}|{\cal I})$ is the degree of belief we
ascribe to the proposition that no gravitational waves from a core
collapse supernova in the Virgo cluster passed through our detector in
the last hour, given only our prior understanding of astrophysics;
similarly, $P(H_{0}|g,{\cal I})$ is the degree of belief we ascribe to
the same proposition, give {\em both} the observations $g$ and our
prior understanding of astrophysics.

To understand how $P(H_{0}|{\cal I})$ and $P(H_{0}|g,{\cal I})$ are
related to each other we need to recall two properties of probability.
The first is unitarity: probability summed over all alternatives is
equal to one. In our example, the two alternatives are, given the
observation $g$, a supernova occurred or it did not:
\begin{equation}
P(H_{0}|g,{\cal I}) + P(\neg{}H_{0}|g,{\cal I}) = 1.
\end{equation}
The second property we need to recall is Bayes Law, which describes
how conditional probabilities
``factor'':
\begin{equation}
P(A|B,C)P(B|C) = P(A,B|C) = P(B|A,C)P(A|C).
\end{equation}
Combining unitarity and Bayes Law it is straightforward to show that
\begin{equation}
P(\neg{}H_{0}|g,{\cal I}) = {\Lambda(g)\over\Lambda(g) +
P(H_{0}|{\cal I})/P(\neg{}H_{0}|{\cal I})}
\label{eq:p(H0|g)}
\end{equation}
where
\begin{eqnarray}
\Lambda(g) &=& {P(g|\neg{}H_{0},{\cal I})/ P(g|{H}_{0},{\cal I})}\\
P(g|H_{0},{\cal I}) &=& \left(\begin{array}{l}
\mbox{probability that $g$ is a sample of}\\
\mbox{detector output when $H_{0}$ is true} 
\end{array}\right)\label{eq:p(g|H0)}\\
P(g|\neg{}H_{0},{\cal I}) &=& \left(\begin{array}{l}
\mbox{probability that $g$ is a sample of}\\
\mbox{detector output when $H_{0}$ is false} 
\end{array}\right)\label{eq:p(g|notH0)}
\end{eqnarray}

The two probabilities $P(g|H_{0},{\cal I})$ and $P(g|\neg{}H_{0},{\cal
  I})$ depend on the statistical properties of the detector noise and
the detector response to the gravitational wave signal.  In some cases
they can be calculated analytically; in other circumstances it may be
necessary to evaluate them using, {\em e.g.,} Monte Carlo numerical
methods.  Regardless of how one approaches data analysis --- as a
Bayesian or as a Frequentist --- the detector must be sufficiently
well characterized that these or equivalent quantities are calculable.

Equation \ref{eq:p(H0|g)} describes how our degree of belief in the
proposition $\neg{}H_{0}$ evolves as we review the observations.  If
$\Lambda$ is large compared to the ratio $P({H}_{0}|{\cal
  I})/P(\neg{}H_{0}|{\cal I})$ then our confidence in $\neg{}H_{0}$
increases; alternatively, if it is small, then our confidence in
$\neg{}H_{0}$ decreases.  If $\Lambda$ is equal to unity --- {\em
  i.e.,} the observation $g$ is equally likely given $H_{0}$ or
$\neg{}H_{0}$ --- then the posterior probability $P(H_0|g,{\cal I})$
is equal to the prior probability $P(H_0|{\cal I})$ and our degree of
belief in $H_0$ is unchanged: we learn nothing from the observation.

More complex hypotheses are analyzed in the same way: if 
\begin{eqnarray}
H_{0} &=& \left(
\mbox{no signal was incident on the detector during the last hour}
\right)\\
H_{{\boldsymbol\theta}} &=& \left(\begin{array}{l}
\mbox{the signal described by ${\boldsymbol\theta}$ was incident}\\
\mbox{on the detector during the last hour}
\end{array}\right),\label{eq:Htheta}
\end{eqnarray}
where ${\boldsymbol\theta}$ is some non-zero set of parameters,
exhaust all possible alternative states of nature, then
\begin{equation}
P({\boldsymbol\theta}|g,{\cal I}) = 
{\Lambda(g|{\boldsymbol\theta})\over
\Lambda + P(H_{0}|{\cal I})/P(\neg{}H_{0}|{\cal I})}
P({\boldsymbol\theta}|\neg{}H_0{\cal I})
\end{equation}
where
\begin{eqnarray}
\Lambda(g|{\boldsymbol\theta}) &=& 
{P(g|H_{{\boldsymbol\theta}},{\cal I}) / P(g|H_{0},{\cal I})}
\label{eq:Lambda(g|theta)}\\
\Lambda(g) &=& \int 
d^n\!\theta\,\Lambda(g|H_{{\boldsymbol\theta}},{\cal
  I})P({\boldsymbol\theta}|\neg{}H_0,{\cal I})\\
P(\neg{}H_{0}|{\cal I}) &=& 1-P(H_{0}|{\cal I}).
\end{eqnarray}
The alternative hypotheses represented by ${\boldsymbol\theta}$ may
represent the radiation from different sources ({\em e.g.,}
supernovae, inspiraling compact binaries, stochastic signal, {\em
  etc.}), different numbers of sources (radiation from more than one
example of a source, or from more than one kind of source), or details
about the particular sources (signal amplitude, source sky position,
inspiraling binary chirp mass, {\em etc.}).

We can now answer the question that began this section. As Bayesians,
we understand confidence to mean {\em degree of belief\/} in the
proposition that radiation originating from a new supernova in the
Virgo cluster was incident on a particular detector during a
particular hour. In response we calculate a quantitative assessment of
our degree of belief in that proposition --- the probability that the
proposition is true.



\section{Guessing Natures State: Frequentist Data
  Analysis}\label{sec:Frequentist} 

With what confidence can we conclude that, in the last hour, the
gravitational waves from a new core collapse supernova in the Virgo
cluster of galaxies passed through our gravitational wave detector?

As before, we have the hypothesis $H_{0}$ and its logical negation, 
$\neg{}H_{0}$.  The gravitational waves from a new Virgo cluster 
supernova either passed through our detector, or they did not.  Our 
goal is to determine, as best we can, which of these two alternatives 
correctly describes what happened.

We decide which alternative is correct by consulting our observation
$g$. Operationally, we adopt a rule or a procedure that, when applied
to $g$, leads us to accept or reject $H_{0}$. The question that began
this section asks us to determine the most reliable rule or procedure.

There are many procedures that we can choose from.  Some are silly:
for example, always rejecting $H_{0}$ is a procedure.  Similarly,
accepting $H_{0}$ if a coin flip comes-up heads is a procedure.  Some
procedures are more sensible: we can calculate a characteristic
amplitude from the observation ({\em e.g.,} a signal-to-noise ratio)
and reject $H_{0}$ if the amplitude exceeds a threshold.  Nature
doesn't speak clearly and, often, some crucial information is hidden
from us; so, no procedure will, in the end, be perfect and every rule
will, on unpredictable occasions, lead us to erroneous conclusions.
Still, some procedures are clearly better than others: the question
is, how do we distinguish between them quantitatively?

Better procedures are those that are less likely to be in error;
consequently, we focus on the error rate of different procedures.  For
our simple problem, where we want to decide only if we have or have
not observed the radiation from a supernova (reject or accept
$H_{0}$), there are two kinds of errors a decision procedure can make:
\begin{enumerate}
\item If no radiation is present ($H_{0}$ true), the rule may
  incorrectly lead us to conclude that radiation is present: 
  a {\em false alarm,} or type I, error.
\item If radiation is present ($H_{0}$ false), the rule may
  incorrectly lead us to conclude that radiation is absent: 
  a {\em false dismissal,} or type II, error.
\end{enumerate}
The false alarm rate is generally denoted $\alpha$ while the 
false dismissal rate is denoted $\beta$.\footnote{When we have a more 
complex set of hypotheses --- for example, when we are asked to choose 
between the set of alternative hypotheses $H_0$ and 
$H_{{\boldsymbol\theta}}$ (cf.\ eq.~\ref{eq:Htheta}) --- there are
additional measures of error: {\em e.g.,} the difference between the 
mean of the estimate of ${\boldsymbol\theta}$ and its actual value, or 
the variance of the estimates, {\em etc.}}

The false alarm and false dismissal rates are {\em frequency 
probabilities.} If we have an ensemble of identical detectors 
simultaneously observing the same system for which $H_{0}$ (or 
$\neg{}H_{0}$) is true, and we apply our rule to each observation, 
then the limiting error rate is given by the false alarm (or 
dismissal) rate.  When, in the last section, we wore our Bayesian hat, 
we understood probability to mean {\em degree of belief;} now, 
however, we understand probability to mean {\em limiting frequency of 
repeatable events.} The distinction in meaning is associated with the 
distinctly different interpretations of the question posed at 
the beginning of this and the last section.

We have seen that, even in the simple case at hand (a single
hypothesis that we must accept or reject), there are at least two
distinct kinds of errors that an inference procedure can make.  Our
measure of a rule's reliability thus involves at least two dimensions,
and may involve more. How, then, do we order the rules to settle upon
a best, or optimal, rule?

To rank rules we must reduce the several error measures that describe 
a procedure's performance to a single figure of merit.  How we choose 
to do this depends on the nature of our problem.  In our case, rules 
that distinguish between $H_{0}$ and $\neg{}H_{0}$ are characterized 
by their false alarm and false dismissal rates; consequently, our 
criteria for ranking rules should depend on relative intolerance to 
false alarms and false dismissals.  For example, if we are testing for 
the presence of antibodies in an effort to diagnose and treat a 
serious illness, we might be very concerned to keep the false 
dismissal rate low, and not nearly as worried about a high false alarm 
rate: after all, a false dismissal might result in death, while a 
false alarm only in an unnecessary treatment with less serious 
repercussions.  Judges or juries in criminal trials faces different 
concerns: false dismissals let criminals go free, while
false alarms send the innocent to prison --- neither alternative 
being very palatable.  Finally, in the case of gravitational wave 
detection, we may (at least initially) be very concerned to avoid 
false alarms, even at the risk of falsely dismissing many real 
signals.

Thus, in order to provide a relative ranking of different inference
procedures for detection or parameter estimation we must construct an
{\em ad hoc\/} figure of merit reflecting the particular nature of the
decision to be made. We term the best rule, under that {\em ad hoc\/}
criteria, the ``optimal'' rule. ``Optimality'', however, is a relative
concept: if the criteria change, the ``optimal'' rule changes also. In
the three examples given above, the criteria might be
\begin{itemize}
\item {\em medical diagnosis:} fix a maximum acceptable false
  dismissal rate and choose the rule that, among all rules whose false
  dismissal rate is so constrained, has the minimum false alarm rate;
\item {\em criminal justice:} choose a rule whose weighted total error
  $\alpha\cos\phi+\beta\sin\phi$ is minimized ($\phi$ being a matter
  of personal choice for an individual judge or juror); 
\item {\em gravitational wave detection:} fix a maximum acceptable
  false alarm rate and choose the rule that, among all rules whose
  false alarm rate is so constrained, has the minimum false dismissal
  rate. 
\end{itemize}

False alarm and dismissal rates describe our confidence in the 
long-run behavior of the associated decision rule.  To understand the 
implications of this measure of confidence, suppose that we have not 
one, but $N$ independent and identical detectors all observing during 
the same hour.  We use the same test, with false alarm rate $\alpha$ 
and false dismissal rate $\beta$, on the observations made at each 
detector, and find that, of these $N$ observations, $m$ lead us 
(through our inference rule) to reject $H_0$ and $N-m$ lead us to 
accept $H_0$.  For a concrete example, suppose $\alpha$ is 1\%,
$N$ is ten and $m$ is three.

The probability of obtaining this outcome when the signal is absent
($H_0$ is true) is the probability of obtaining $m$ false alarms in
$N$ trials, or
\begin{equation}
P(m|H_0,N) = {N!\over(N-m)!m!}\alpha^m(1-\alpha)^{N-m}.
\end{equation}
In our example, $P(m|H_0,N)$ evaluates to $1.1\times10^{-4}$. It is
thus very unlikely that we would have made this observation if the
signal were absent. Does this mean we should conclude the signal is
present with, say, $99.99$\% confidence?

No. $P(m|H_0,N)$ describes the probability of observing $m$ false
alarms out of $N$ observations.  When the signal is {\em present,}
however ({\em i.e.,} when $H_0$ is false), there are {\em no\/} false
alarms and both $\alpha$ and $P(m|H_0,N)$ are irrelevant. There are,
however, $N-m$ false dismissals; thus, the relevant quantity is
$P(m|\neg{}H_0,N)$, the probability of observing $N-m$ false {\em
  dismissals:}
\begin{equation}
P(m|\neg{}H_0,N) = {N!\over(N-m)!m!}(1-\beta)^m\beta^{N-m}.
\end{equation}
If, in our example, the false dismissal rate $\beta$ is 10\%, then the
probability of observing seven false dismissals out of ten trials is
is $8.7\times10^{-5}$.

The particular outcome of our example --- three positive results out 
of ten trials --- may be, overall, unlikely; however, it is more 
unlikely to have occurred when the signal is present then when it is 
absent.  Despite the apparently overwhelming improbability of three 
false alarms in ten trials, it is nevertheless, slightly more likely 
than the alternative of seven false dismissals in ten trials.


We can now answer the question that began this section.  As 
Frequentists, we understand the question to ask for the overall error 
rates of the best general procedure for deciding between the 
alternative hypotheses.  We thus calculate the error rates for 
different inference rules, choose appropriate criteria for ranking the 
different rules, and find the best rule and its corresponding error 
rates.

Contrast this with our understanding of the identically worded 
question posed to us as Bayesians.  As Bayesians, we understood 
confidence to mean the degree of belief that we should ascribe to 
alternative hypotheses; as Frequentists, we understand confidence to 
refer to the overall reliability of our inference procedure.  As 
Bayesians we responded with a quantitative assessment of our degree of 
belief in the alternative hypotheses, {\em given a particular 
observation made in a particular detector over a particular period of 
time;} as Frequentists, we responded with an assessment of the 
relative frequency with which our rule errs given each alternative 
hypothesis.  As Bayesians, we did not make a choice between alternative 
hypotheses; rather, we rated them as more or less likely to be true in 
the face of a particular observation.  As Frequentists, on the other 
hand, we do make choices and our concern is with the error 
rate of our procedure for choosing, averaged over many different 
observations and many different decisions.

Frequentist analyses have particular utility when it is possible to
make repeated observations on identical systems: {\em e.g.,} particle
collisions in an accelerator, where each interaction of particle
bunches is an ``experiment.'' Bayesian analyses, on the other hand,
are particularly appropriate when the observations or experiments are
non-repeatable: {\em e.g.,} when the sources are, like supernovae,
non-identical and destroy themselves in the process of creating the
signal. In the latter case, we are generally particularly interested
in the properties of the individual systems and would prefer a measure
of the relative degree of belief that we should ascribe to, for
example, the proposition that the signal originated from different
points in the sky.

\section{Example: Data Analysis for Stochastic
  Signals}\label{sec:example}

As an example of how Bayesian and Frequentist statistical 
methodologies lead to quantitatively different analyses, consider how 
one might search for a stochastic gravitational wave signal, making 
use of observations in a pair of gravitational wave detectors.  Over 
the last several years a Frequentist analysis of this problem has been 
developed~\cite{michelson87a,christensen92a,flanagan93a,allen97c}.  In 
\S\ref{sec:FreqApp} I outline that analysis (see also Allen's 
contribution to these proceedings).  In 
\S\ref{sec:BayesApp} I outline an alternative, Bayesian analysis of 
the same problem~\cite{finn97b}.

\subsection{A Frequentist approach}\label{sec:FreqApp}

Very briefly, the Frequentist analysis developed in the
literature~\cite{michelson87a,christensen92a,flanagan93a,allen97c}
begins with the observation that a stochastic signal will lead to a
correlated response in the output of physically distinct detectors,
while instrument noise in the same detectors is likely to be largely
uncorrelated.  This observation leads us to consider Frequentist
decision rules based on a correlation of the output of several
detectors.  Focus attention on two detectors, whose discretely sampled
output over the interval $[0,T)$ are given by the sequences $h_1[k]$
and $h_2[k]$, with $k$ running from $0$ to $N_T-1$.  Focus attention
on the generalized correlation
\begin{equation}
\mu_Q(h_1,h_2) =
\sum_k^{N_T}\sum_j^{N_T} h_1[k]Q[k-j]h_2[j].
\label{eq:qtest} 
\end{equation}
The coefficients $Q[k-j]$ are, for now, arbitrary. 

If no signal is present, then the distribution of $\mu_Q$ depends
on the statistical properties of the detector noise; for instance, if
the noise is uncorrelated between the detectors, then $\mu_Q$ will
have vanishing mean. If the signal is present, the distribution of
$\mu_Q$ depends on statistical properties of both the noise and
the stochastic signal and will, in general, be different from the case
of no signal because the signal leads to a correlation in the output
of the two detectors.  Let $\sigma^2_Q$ be the variance of $\mu_Q$
{\em in the absence of a signal,} {\em i.e.,}
\begin{equation}
\sigma^2_Q \equiv \overline{\mu^2_Q}
\end{equation}
where the average is over different instantiations of the random 
detector outputs when these are detector noise alone.  Define also the 
{\em signal-to-noise ratio\/} $\rho_Q$:
\begin{equation}
\rho_Q(h_1,h_2) \equiv \mu_Q(h_1,h_2)/\sigma_Q.\label{eq:rhoQ}
\end{equation}
If no signal is present and the noise in the two detectors is 
uncorrelated, $\rho_Q$ will, averaged over many trials, vanish.  On 
the other hand, if there is a signal present then $\rho_Q$ will, over 
many trials, have a non-zero average value which is proportional to 
the signal strength and independent of the normalization of $Q$.  A 
Frequentist decision rule is to fix $Q$ and evaluate $\rho_Q$, 
deciding that a signal is present if $\rho_Q$ exceeds a fixed 
threshold.  The amplitude of $\mu_Q$ will, on average and for 
sufficiently large signal amplitudes, be proportional to the squared 
signal amplitude.

The freedom in the choice of $Q$ can be used to tune the test. Suppose
that the spectrum of the stochastic signal is known up to a constant,
unknown amplitude. Assume that, for fixed non-zero signal strength,
better tests give larger $\overline{\rho_Q}$; then, knowing the
statistical properties of the signal and the noise we can maximize
$\overline{\rho_Q}$ over the coefficients $Q$ to find the ``best''
test of the form given in equation \ref{eq:qtest} (best in this case
meaning the test that gives greatest signal-to-noise for fixed
signal).  Since $\overline{\rho_Q}$ is proportional to the signal
amplitude, if we decide that a signal is present, $\rho_Q$ also
provides an estimate of the signal amplitude.

\subsection{A Bayesian approach}\label{sec:BayesApp}

Observations enter Bayesian analyses through the likelihood function
(cf.\ eq.~\ref{eq:Lambda(g|theta)}). Regard the discretely sampled
detector outputs $h_1[k]$ and $h_2[k]$ as components of ${\bold h}[k]$,
the vector-valued output of a gravitational wave receiver consisting
of two detectors. Assume for simplicity that the noise and signal both
have Gaussian-stationary statistics (this is the case for the usual
stochastic signals considered and is also assumed in the Frequentist
analyses to date~\cite{christensen92a,flanagan93a,allen97c}). The
statistical properties of the receiver noise ${\bold n}[k]$ are then
characterized fully by the sequence of noise correlation matrices
${\bold C}[k]$:
\begin{eqnarray}
{\bold C}[j-k] &=& \overline{{\bold n}[j]\otimes{\bold n}[k]}\\
\left|\left|{\bold C}[j-k]\right|\right|
&=&\left|\left|\begin{array}{cc}
\overline{n_1[j]n_1[k]}&\overline{n_1[j]n_2[k]}\\
&\\
\overline{n_2[j]n_1[k]}&\overline{n_2[j]n_2[k]}
\end{array}\right|\right|.
\end{eqnarray}
The probability that an observation ${\bold h}$ is a sample of receiver
noise is a multi-variate Gaussian,
\begin{eqnarray}
P({\bold h}|0) &=& {
\exp\left(-\left<{\bold h},{\bold h}\right>_{{\bold C}}\right)/
2\pi\sqrt{\det\left|\left|{\bold C}\right|\right|}}\\
\left<{\bold h},{\bold h}\right>_{{\bold C}} &=&
{1\over2}
\sum_{j,k}^{N_T}
{\bold h}[j]\cdot\left|\left|{\bold C}\right|\right|^{-1}_{jk}
\cdot{\bold h}[k],
\end{eqnarray}
where the covariance matrix $||{\bold C}||^{-1}$ is the inverse of the
$2(2N_T-1)\times2(2N_T-1)$ block matrix $||{\bold C}||$ composed from
the sequence ${\bold C}[k]$: $\left|\left|{\bold C}\right|\right|_{jk} =
\left|\left|{\bold C}[j-k]\right|\right|$.

In the presence of a Gaussian-stationary stochastic signal the
receiver output remains Gaussian-stationary; however, the covariance
changes. The new covariance ${\bold K}$ depends on the
auto-correlation of the receiver noise ${\bold C}$ and the
auto-correlation of the receiver response to the stochastic signal
${\bold S}({\boldsymbol\theta})$, which is characterized by
${\boldsymbol\theta}$. Thus, the probability that observed receiver
output ${\bold h}$ is a sample of noise plus signal is
\begin{eqnarray}
P({\bold h}|{\boldsymbol\theta}) &=& {
\exp\left(
-\left<{\bold h},{\bold h}\right>_{{\bold K}({\boldsymbol\theta})}
\right)/
2\pi\sqrt{\det\left|\left|{\bold K}({\boldsymbol\theta})\right|\right|}
}\\
{\bold K}({\boldsymbol\theta}) &=& {\bold C} +
{\bold S}({\boldsymbol\theta}). 
\end{eqnarray}

The likelihood function is the ratio of
$P({\bold h}|{\boldsymbol\theta},{\cal I})$ to  
$P({\bold h}|0,{\cal I})$, or 
\begin{equation}
\Lambda({\bold h}|{\boldsymbol\theta}) = \sqrt{
\det\left|\left|{\bold C}\right|\right|\over
\det\left|\left|{\bold K}({\boldsymbol\theta})\right|\right|
}\,{
\exp\left(\left<{\bold h},{\bold h}\right>_{{\bold C}}\right)\over
\exp\left(\left<{\bold h},
{\bold h}\right>_{{\bold K}({\boldsymbol\theta})}\right)
}
\end{equation}

The optimal test described in \S\ref{sec:FreqApp} is based on the test
statistic $\rho$, which is $\rho_Q$ (cf.\ eq.~\ref{eq:rhoQ}) with $Q$
chosen to give the largest possible average value for $\rho_Q$ in the
presence of a signal. This statistic is, by construction, linear in
the output of each of the two detectors. The likelihood function, on
the other hand, has a more complex dependence on the detector outputs;
in particular, $\Lambda(g|\theta)$ {\em is not a function of the test
  statistic $\rho$.} As a result, conclusions reached through the
Frequentist test summarized in \S\ref{sec:FreqApp} will be
quantitatively different than those reached in a Bayesian analysis.

A difference does not imply that one analysis is right and the other
wrong. As we have seen, Bayesian and Frequentist analyses do not
address the same questions; so, they are not required to reach
``identical'' conclusions. On the other hand, it may well be that one
analysis is more appropriate or responsive to our concerns than the
other. We can only make the choice of appropriate analysis tool when
we understand the distinction between them.

\section{Conclusions}

Bayesian data analyses address how observations affect our {\em degree
  of belief\/} in propositions about nature: for example, in the
proposition that radiation originating from a new supernova in the
Virgo cluster was incident on a particular gravitational wave detector
during a particular hour. The result of a Bayesian analysis is a
quantitative measure of our degree of belief in the proposition in the
face of an observation --- the probability that the proposition is
true.

Frequentist data analyses address our confidence in a general
procedure for deciding, {\em e.g.,} whether a signal is present or
absent.  Frequentists do not assess degree of belief or confidence in
any particular conclusion; rather they assess confidence in the
procedure used for reaching conclusions.  The result of a Frequentist
analysis is a conclusion about the state of nature, made by a
carefully chosen rule, upon examination of the data.  {\em
  Confidence\/} in that conclusion is an assessment of the rule's {\em
  average performance:} if one were to observe the same source in the
same detector many times, how frequently, or by what average amount,
would our procedure for deciding err?

In a Frequentist analysis, confidence is the reliability of a
procedure; in a Bayesian analysis, it is the degree of belief ascribed
to a hypothesis.  Since astrophysical events that generate
gravitational waves tend to be unique, gravitational wave data
analysis that focuses on the properties of individual sources should
be Bayesian.  When studying individual sources we are very interested
in quantifying our uncertainty (in, {\em e.g.,} the source chirp mass)
given the only observations we have.  We are generally not interested
in knowing how well some procedure might do in estimating, {\em e.g.,}
the chirp mass if we could observe the same source many times.  We
only get one look at most sources --- stochastic background and
periodic sources being the exceptions --- and Bayesian techniques are
more suited to the study of individual sources.

Even in the case of stochastic and periodic signals, Bayesian analyses
address questions closer to our immediate interests.  Given a long
enough observation, Frequentist and Bayesian analyses converge to
equivalent conclusions: in the Bayesian analysis, the long observation
leads us to increasingly narrow the degree of our uncertainty, while
in a Frequentist analysis, the long series of observations leads to a
series of conclusions (from repeated application of our inference
rule) whose distribution will be consistent with only one possible
hypothesis about nature.  In the Bayesian analysis, however, we know
at any point along the way what our degree of uncertainty is, while in
the Frequentist analysis our conclusions are not about our degree of
uncertainty (in, {\em e.g.,} the presence or absence of a signal) but
about how likely it is that one can distinguish between the
alternatives ({\em e.g.,} signal absent/present, or signal amplitude)
given the number of different observations made up to that point.

\section*{Acknowledgments}

It is a pleasure to thank the organizers and CERN for their efforts in
putting together a very successful meeting. I am also grateful for
the hospitality of the LIGO Project and the California Institute of
Technology, where this manuscript was prepared.  This work was
supported by National Science Foundation awards PHY 93-08728 and PHY
95-03084.

\section*{References}

\end{document}